# Vortex ratchet reversal: The role of interstitial vortices.

D. Perez de Lara[1,3], M. Erekhinsky[2], E. M. Gonzalez[3], Y. J. Rosen[2], Ivan K. Schuller[2], J. L. Vicent[1,3]

[1] IMDEA-Nanociencia, Cantoblanco, 28049 Madrid, Spain

[2] Physics Department, University of California-San Diego, La Jolla, California 92093, USA

[3] Departamento Fisica de Materiales, Universidad Complutense, 28040 Madrid, Spain.

**Abstract**

Triangular arrays of Ni nanotriangles embedded in superconducting Nb films exhibit unexpected dynamical vortex effects. Collective pinning with a vortex lattice configuration different from the expected fundamental triangular "Abrikosov state" is found. The vortex motion which prevails against the triangular periodic potential is produced by channelling effects between triangles. Interstitial vortices coexisting with pinned vortices in this asymmetric potential, lead to ratchet reversal, i.e. a DC output voltage which changes sign with the amplitude of an applied alternating drive current. In this landscape, ratchet reversal is always observed at all magnetic fields (all numbers of vortices) and at different temperatures. The ratchet reversal is unambiguously connected to the presence of two locations for the vortices: interstitial and above the artificial pinning sites.

**INTRODUCTION**

Physics equations exhibit time reversal symmetry, however many physical, chemical and biological systems are not symmetric in time. Thus studies of simple physical systems in which time reversal asymmetry can be engineered in a simple and reproducible way, are worthy of research since they provide insight into the origin of the physics of irreversibility. An interesting way in which irreversibility may appear is when the driving force is periodic but the response is unidirectional ("ratchet"). Moreover, an interesting and unique effect is the ratchet reversal (the change in sign of the unidirectional response) as a function of relevant parameters. Recently, ratchet reversal has been achieved in optical [1], Josephson junction [2, 3] and superconducting film based systems [4], and therefore this seems to be a very general phenomenon [5-7]. Vortex ratchet reversal is due to collective effects such as deformations of the vortex lattice, the appearance of interstitial vortices in an effective pinning potential created by the pinned vortices, or the creation of interstitial or vacancy sites in ordered commensurate vortex configurations [8-11]. Some of the first results predicting ratchet reversals in interacting particle systems were proposed by [8].

One simple system which exhibits time irreversibility, and can be produced and studied in a systematic way, consists of an array of asymmetric magnetic pinning sites in proximity to a superconducting film. In this system, the application of an AC current can produce a DC voltage ("ratchet effect") which; i) depends in interesting ways on the various system parameters (geometry, magnetic field, temperature etc.) and ii) exhibits sign reversals as a function of several important parameters of the system. This ratchet effect is caused by the motion of superconducting vortices in an asymmetric potential subject to an external alternating driving force. The array may have an intrinsic asymmetry built using symmetric individual pinning sites [12] or the individual pinning sites may be asymmetric although the array is symmetric [4]. The important parameters which control the superconductivity are the coherence length and penetration depth and therefore the scale of the physics is set by these two physical parameters. The pinning sites may be magnetic or non-magnetic and therefore pinning may arise from structural [4] and/or magnetic [13] effects. Experimentally it is customary to measure the rectified DC voltage when the system is subject to an alternating drive current. Therefore the DC voltage, $V_{DC}$, which is a measure of the average velocity $\langle \vec{v} \rangle$ of the vortex lattice depends on the: a) externally applied perpendicular H field, which determines the number of vortices, b) alternating current $J_{AC} = I_{AC} \sin(\omega t)$ related to the driving force F on the vortex lattice, ($I_{AC}$ the current amplitude and $\omega$ the frequency) and c) the pinning potential. Naively the motion of the vortices is expected along the Lorentz force, which is perpendicular to the current direction. However, due to "channelling" effects arising from the strong pinning by the magnetic pinning landscape, the vortices move locally along directions which are not parallel to the Lorentz force. Of course globally the average vortex motion is along the Lorentz force. One of the interesting effects in this system is the so called "ratchet reversal" in which there is a sign change of the rectified DC voltage [4].

In order to understand the origins and implications of the ratchet effect it is important to study a well-defined system where the parameters are systematically varied. The ratchet effect in superconducting films was originally observed in a square array of triangular magnetic pinning sites as a function of $I_{AC}$ [4]. Ratchet reversal as a function of $I_{AC}$ however was observed only above a threshold H field corresponding to a number of vortices per array unit cell greater than 3. Several theoretical models were advanced to explain the origin of ratchet reversal, some propose the presence and independent motion of interstitial vortices [4], and others require the reorganization of the whole vortex lattice and its collective motion [14]. Here we compare our studies of a triangular array of pinning sites to the earlier studied square array. The individual (triangular) pinning sizes relative to the superconducting parameters (coherence length and penetration depth) are maintained which allows us to distinguish between different classes of theoretical models.

**EXPERIMENTAL METHODS**

Here we studied several superconducting Nb films covering arrays of magnetic Ni triangles on Si (100) substrates. The equilateral triangles with sides close to 600 nm and thickness 40 nm were arranged in a triangular array with a periodicity of around 700 nm. Fig. 1 shows a SEM picture of the array. The Ni triangles were prepared by electron beam lithography using polymethyl methacrylate resist and lift-off. The Ni was deposited by electron beam evaporation in a system with a base pressure of $10^{-7}$ Torr. The 100 nm thick Nb film was deposited using magnetron sputtering with a base pressure of $10^{-8}$ Torr above the nanostructured Ni array. Electrical leads were patterned using photolithography and etching to form a 40μm × 40μm bridge, which allows propagating currents and measuring voltage drops in two perpendicular directions.

The electrical resistivity of the hybrid system was measured using the standard 4-point probe method, with a magnetic field applied perpendicular to the sample plane. Using this geometry we are able to induce in-plane vortex motion parallel and perpendicular to the symmetry axes of the nanotriangles. This magnetoresistance for several similar samples was obtained at temperatures close to the critical temperature in a liquid Helium cryostat with a superconducting solenoid and a variable temperature insert. The superconducting critical temperature of the devices is 8.6 K, the penetration depth and the coherence length at $0.99T/T_c$ are 1.5 μm and 97 nm respectively, and $\lambda$=298 nm and $\xi$=9.7 nm.

An alternating current ($\vec{J}_{AC}$) injected perpendicular to the triangle symmetry axis induces an alternating Lorentz force on each vortex $\vec{F}_L = \vec{J}_{AC} \times \vec{z}\phi_0$ ($\phi_0$ is the quantum fluxoid and $\vec{z}$ is a unit vector parallel to the applied magnetic field $\vec{B}$). Although the time-averaged force on the vortices is zero $\langle F_L \rangle = 0$, in the presence of an asymmetric potential, a non-zero DC voltage ($V_{DC}$) can develop. This is the so called ratchet effect. This voltage is proportional to the time averaged velocity ($\langle \vec{v} \rangle$) of the vortex lattice. Positive voltage here corresponds to the vortices moving from base to the tip of the triangular pinning sites (positive direction) and the negative voltage appears when vortices move from tip to base (negative direction). The ratchet effect measurements were performed at the highest attainable frequency for our experiments (10 kHz).

**RESULTS**

Close to the superconducting critical temperature, the magnetoresistance of superconducting thin films with periodic arrays of pinning centers exhibits minima for fields corresponding to an integer number of vortices per plaquette [15, 16]. Fig. 2 shows the magnetoresistance with the current applied parallel and perpendicular to the symmetry axis of the triangles, corresponding to vortex lattice motion perpendicular and parallel to the triangle symmetry axis, respectively. In both cases resistance minima appear at $\Delta H = 36$ Oe (see inset, Fig. 2). However, the theoretical matching field corresponding to a triangular unit cell 700 nm side is 46 Oe. This is clearly different from the experimental result with a discrepancy of 21% and outside experimental error, magnetic field resolutions of 1 Oe. This implies an experimental vortex density lower than the theoretical density corresponding to one vortex per triangle and therefore a larger vortex lattice area than the triangular lattice. Since the $\Delta H$ is the same for both current directions, the vortex lattice unit cell area is independent of the vortex motion direction (see Fig. 2). Therefore, whatever the detailed vortex lattice geometric arrangement is, under the assumption that the vortex lattice is regular and uniform, interstitial vortices must always be present.

Note also that in all earlier cases including reference 4 the largest disagreement between the calculated and measured matching field was 8% (for instance 35 Oe calculated and 32 Oe measured in ref 4, for a 770 nm x 750 nm array). Therefore we believe that the discrepancy observed here is physically significant.

Fig. 3 shows the ratchet voltage at: (a) three different temperatures ($T/T_c = 0.99$, 0.98 and 0.97) and (b) several matching fields corresponding to integer numbers of the first matching field between the vortex lattice and the pinning array. Below a (field and temperature dependent) threshold for low AC drive amplitudes, no ratchet effect is observed. As the drive amplitude increases a negative DC voltage develops and at a temperature dependent drive amplitude the sign of the DC voltage switches to positive. Therefore, when the Lorentz force is large enough to set the weakly bound interstitial vortices in motion, they move in the negative direction (negative DC voltage). This is due to an "inverted" potential produced by the triangular shaped empty areas pointing in the direction opposite to the Ni triangles. Vortices pinned on the triangles need a higher driving force to be set in motion (positive DC). These vortices require less force to exit the Ni-triangle tips than the Ni-triangles base, resulting in a positive DC ratchet. The driving force (or current) required to reverse the ratchet effect is related to the pinning potential strength and is approximately the same for different fields, at constant temperature (slightly above 4 mA at T=0.97 $T_c$, see Fig 3(b)). When the Lorentz force is significantly larger than the pinning force, the vortex-lattice time averaged velocity approaches zero ($\langle \vec{v} \rangle = 0$). Ratchet reversal with drive amplitude and a variable number of particles is very unusual in most structurally asymmetric systems. To the best of our knowledge, such an unusual phenomenon in a simple electronic system has been reported only once [16].

The DC rectified amplitude develops in a very systematic way as a function of temperature and field in this kind of ratchet, denoted as a "rocking" ratchet. Fig. 4 shows the evolution of the ratchet reversal effect with various parameters which characterize it: minimum negative voltage (red down triangles) and maximum positive voltage (black up triangles) highlighted by arrows in the inset of Fig. 4. The minimum voltage increases (becomes more negative) with increasing N whereas the maximum voltage increases up to a peak and then decreases to zero.

**DISCUSSION**

The results presented in Fig. 2 imply that the order in the vortex lattice at the matching fields is identical although the Lorentz force is applied in two structurally asymmetric directions. In contrast, assuming the interstitial picture for the ratchet effect is correct, the motion of the vortices must be very different for the ratchet configuration. Naively one would assume there would be no ratchet for the triangular array. If the vortices moved only along the symmetry axis of pinning-triangles, the interstitial vortices would move in and out of the pinning triangles and vice versa. This would imply the absence of ratchet reversal. Therefore the presence of ratchet reversal implies that vortices, on average moving along the symmetry axis, must be travelling in a zig-zag path from interstitial to interstitial. This type of channelling has been seen elsewhere [17, 18, 19].

We emphasize that, the ratchet reversal is well established experimentally in superconducting films with arrays of structurally asymmetric pinning sites but its origin is very controversial. There are different models used to explain the presence of this reversal. The simplest relies on the number of vortices; the threshold to obtain ratchet reversal is reached once the vortex array density is sufficiently high to produce interstitial vortices in a square array of triangles. Therefore, the reversal should only appear above a critical value of the magnetic field when interstitial vortices are present [4]. The key idea in this model is that the ratchet and reversal are produced by independent vortex motion. Therefore, the reversal indicates that when there is ratchet reversal the vortex lattice breaks into two subsystems moving against each other. An alternative model [14] is based on the idea that with increasing magnetic field the vortex-vortex interaction dominates over the vortex pinning-site interaction which produces a vortex lattice rotation and reconfiguration. As a consequence the collective motion of the whole vortex lattice changes sign. In other models, ratchet reversal is accomplished by increasing the number of vortices as well as vortex lattice disorder [20]. Lattice disorder and strong vortex-vortex interaction lead to multiple ratchet reversals in samples with a pinning potential period similar or smaller than the superconducting penetration length [21]. Additionally, very subtle mechanisms such as two-dimensional instabilities in the ground state of the vortex lattice in competition with the array pinning strengths [22] may induce vortex ratchet reversal as well. In summary, different models, summarized in Table 1, have been reported to explain these interesting vortex ratchet reversals.

The results presented here (Fig. 3 & 4) show very simple and straightforward trends. The system used here is based on the simplest possible array (triangular) of structural ratchet potentials which mimic the symmetry of the Abrikosov vortex lattice ground state. The experiments show ratchet reversal even at the first matching field and when increasing the applied magnetic field the reversal only vanishes at 770 Oe (N=20), very close to the 800 Oe (at T=0.97Tc) upper critical magnetic field (see Fig. 3(b)). The reversed ratchet is controlled by the external driving force and the reversal appears in the whole temperature range close to the critical temperature, where the periodic artificial pinning overcomes the random intrinsic pinning of the superconducting films, i.e. in the experimental temperature window where matching effects appear. The vortex motion is channelled in between the pinning triangles and ratchet reversal is connected with the presence of interstitial vortices. These results agree with the model based on the presence of interstitial vortices without any need of alternative models to explain the onset of ratchet reversal.

The systematic changes as a function of temperature (Fig. 3(a)), drive amplitude, and magnetic field (Fig. 3(b)) hold further clues, improve the understanding and description of the ratchet reversal and allow comparison with other experiments in this field. First, the ratchet signal is only observed if the vortices move along the triangle asymmetry direction, which implies that this effect is connected with the geometrical asymmetry. For all fields and temperatures below a threshold current there is no ratchet effect. With increasing amplitude a negative ratchet appears which implies that small forces set weakly bonded vortices into motion. The decrease of threshold current with increasing field is due to an increase in N which implies that the vortex-vortex interaction increases and therefore the individual interstitial vortices become less and less bound. At this point, the binding energy is smaller, requiring less force to move them. On the other hand, with decreasing temperature the magnitude of the ratchet signal increases and the onset driving current amplitude shifts towards higher drive amplitudes since the binding energy becomes larger.

The field dependence shown in Fig. 4 gives further clues regarding the ratchet effect and establishes noteworthy differences with other experiments [4, 18, 21]. With increasing magnetic field (i.e. N) the ratchet amplitude increases considerably for both ratchet signs (i.e. both types of vortices). The positive ratchet associated with vortices pinned by a magnetic site reaches maximum amplitude and then decreases further with increasing N. This implies that the signal arising from the negative ratchet dominates over the positive one because more vortices are packed interstitially and because the pinning of interstitial vortices becomes weaker as their density increases.

There are important differences between the ratchet effects presented here and other published work [18, 21]. In our system the sign of $V_{DC}$ always changes with the alternating input driving current for any number of vortices (apart from N close to the corresponding critical magnetic field). We emphasize that this particular structurally asymmetric ratchet exhibits a sign reversal in which the DC output voltage polarity can be tuned simply by increasing the amplitude of the alternating drive current.

**CONCLUSIONS**

The results presented here are quite surprising and contrary to naïve expectations. The collective pinning and the ratchet reversal imply that interstitial vortices play a major role through vortex channelling between triangles. We find that rectification of an alternating current exhibits reversal as a function of drive current for all values of the external field (except near the critical field). In contrast, square arrays of triangular pinning sites only exhibit the ratchet effect above a certain minimum driving current and above a critical field. In the square array ratchet reversal occurs only above the third matching field because interstitial vortices only appear above this field. For the triangular arrays, there is no minimum threshold field for the appearance of ratchet reversal because interstitial vortices are always present even at the first matching field.

Research at UCSD funded by the US-NSF and support at UCM was provided by Spanish Ministerio Ciencia e Innovación grants FIS2008-06249 (Grupo Consolidado), Consolider CSD2007-00010, CAM grant S2009/MAT-1726 and Santander-UCM grant GR35/10-A, IMDEA-Nanoscience. We thank Prof. S. Bar-Ad and Y. Bruynseraede for useful conversations.

Figure Captions

Figure 1. Scanning Electron Microscope picture of the triangular array of Ni triangles.

Figure 2. (Color online) Magnetoresistance at $T/T_c=0.99$ of a superconducting Nb thin film on an array of Ni nanotriangles. The superconducting critical temperature was 8.6 K. Linear fit of the matching fields shown in the inset gives ΔH=36 Oe with a linear correlation coefficient of 0.9996, when the current is parallel (red filled circles) and perpendicular (black open circles) to the symmetry axes of the triangles.

Figure 3. (Color online) Ratchet effect in Nb film with array of Ni triangular pinning sites (a) at different temperatures: $T/Tc=0.99$ (black filled circles), 0.98 (red open circles) and 0.97 (blue half-filled circles) and applied field which corresponds to N=1 vortices per unit cell.
(b) at $T=0.97Tc$ for different magnetic fields: N=8 (black full squares), N=12 (red half-filled circles) and N=20 (purple full circles).
The injected alternating currents are parallel to the triangular base. Green crossed circles curve in (a) corresponds to alternating currents perpendicular to the triangular base at $T=0.98Tc$.

Figure 4. (Color online) Dependence of the maximum (black full triangles) and minimum (red full inverted triangles) DC voltages of the ratchet effect on magnetic field at $T=0.97Tc$. Inset shows how the maximum and minimum voltage values are defined.

Table 1. Summary of the different models explaining the origin of ratchet reversal in several superconducting systems.

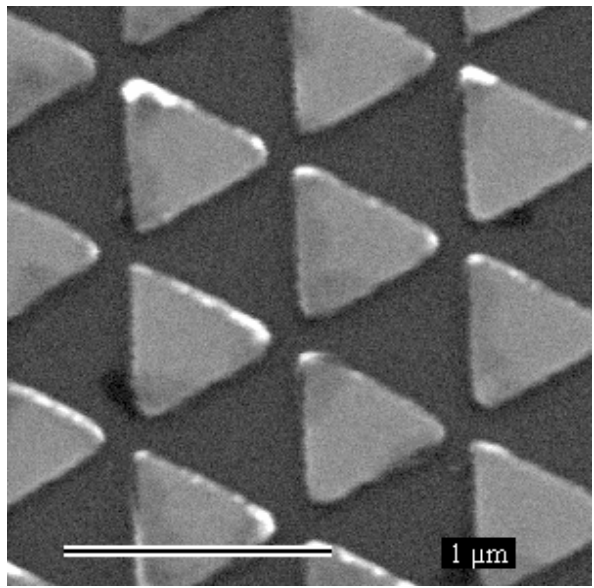

Figure 1

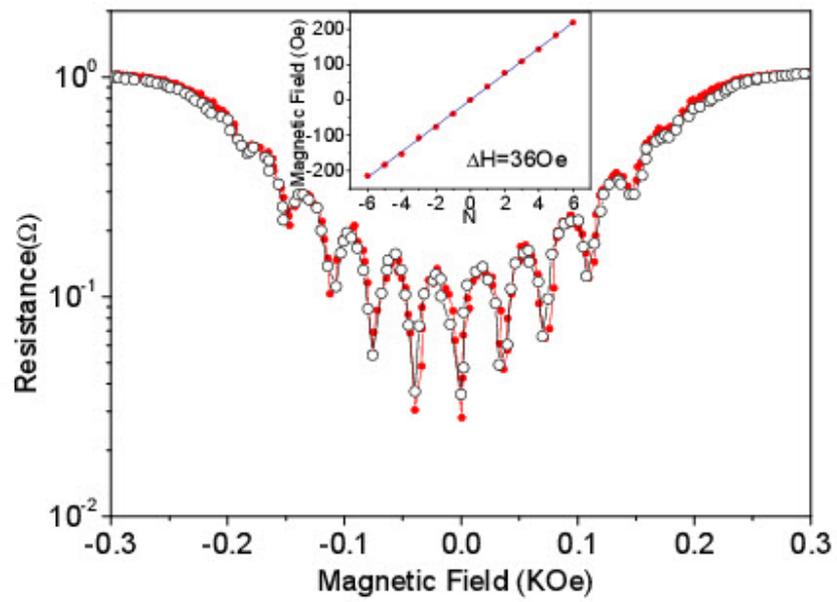

Figure 2

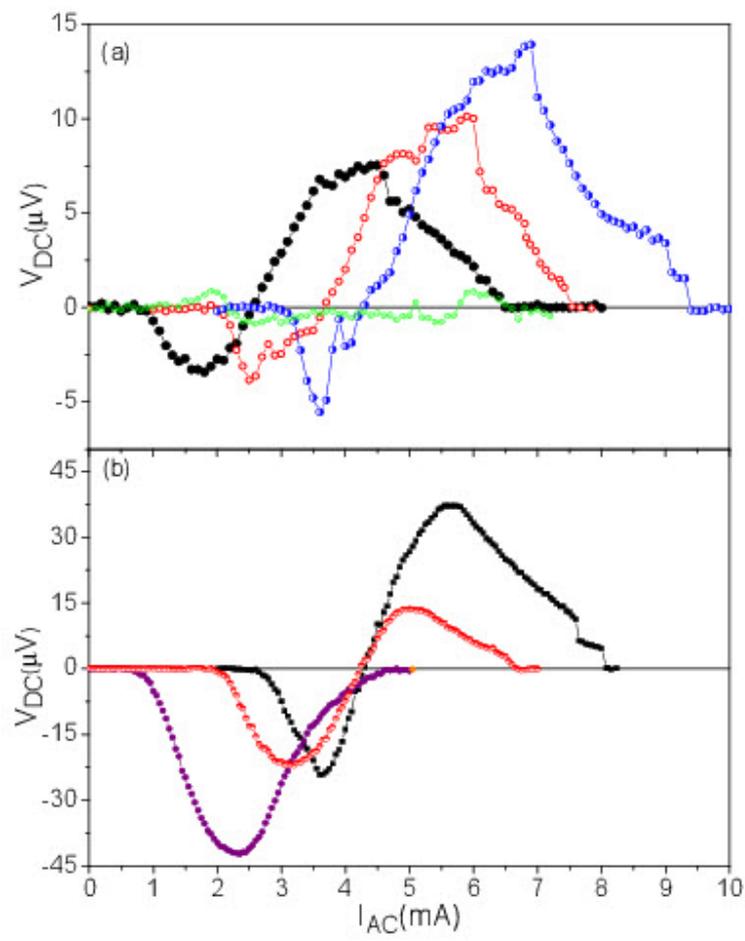

Figure 3

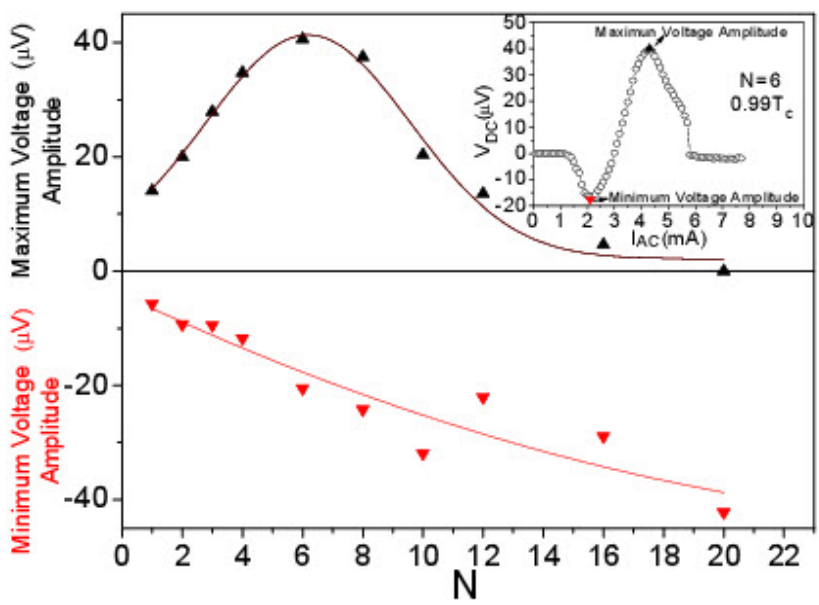

Figure 4

| Ratchet Reversal Origin | Sketch of pinning centers | References |
|---|---|---|
| Number of vortices & Competition between two opposite pinning potentials | 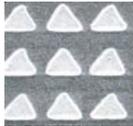 | [*Science* **302**, 1188 (2003)] |
| Instability of the vortex lattice ground state | | [*Phys. Rev.*B**76**, 212507 (2007)] |
| Reconfiguration of the vortex lattice | | [*New Jour. Phys.* **9,** 366 (2007)] |
| Positional disorder and strong Vortex-Vortex interaction | 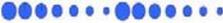 | [*Phys. Rev. Lett.* **99,** 247002 (2007)]<br>[*Phys. Rev. B* **75,** 054502 (2007)] |
| Pinning potential strength & Number of vortices | 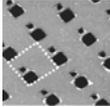 | [*Nature* **440**, 651, (2006)] |

Table 1